\documentclass[aps,prl,twocolumn,showpacs]{revtex4-1}
\usepackage{bm}
\usepackage{amsmath,amssymb}
\usepackage[]{graphicx,color,epstopdf}

\newcommand{\mb}{\mathbf}

\newcommand{\overbar}[1]{\mkern 1.5mu\overline{\mkern-1.5mu#1\mkern-1.5mu}\mkern 1.5mu}

\begin{document}

\title{Rhythmicity, Recurrence, and Recovery of Flagellar Beating}
 
\author{Kirsty Y. Wan and Raymond E. Goldstein}
\affiliation{Department of Applied Mathematics and Theoretical
Physics, Centre for Mathematical Sciences,\\ University of Cambridge, Wilberforce Road, Cambridge CB3 0WA, United Kingdom}
\date{\today}

\begin{abstract}
The eukaryotic flagellum beats with apparently unfailing periodicity, 
yet responds rapidly to stimuli.  Like the human heartbeat, flagellar oscillations are 
now known to be noisy.
Using the alga \textit{C. reinhardtii}, we explore three aspects of nonuniform flagellar beating.  We report the existence of
rhythmicity, waveform noise peaking at transitions between power and recovery strokes, and  
fluctuations of interbeat intervals that are correlated and even recurrent, with memory extending to hundreds of beats. 
These features are altered qualitatively by physiological perturbations.
Further, we quantify the recovery of periodic breastroke beating from transient hydrodynamic forcing.
These results will help constrain microscopic theories on the origins and regulation of flagellar beating.
\end{abstract}

\pacs{47.63.Gd 05.45.-a 87.16.Qp 87.18.Tt}
\maketitle

Patterns of coordinated movement in living organisms, such as walking, running, and galloping, may be variable yet 
simultaneously stable. Such repetitive dynamics are distinguished by their reproducibility, long-time sustainability, and 
robustness to moderate perturbations. In the precise, rhythmic beating of 
the flagella of the alga 
\textit{Chlamydomonas} we find remarkable living oscillators that fulfil these three criteria.
The synchronous beating of its twin $\sim\!\!10$ $\mu$m long flagella allows \textit{Chlamydomonas} to swim a fast 
breaststroke \cite{Racey1981}.
At $\sim\!\!60$ Hz, its flagellar oscillations are \textit{self-sustained} -- repeated mechano-chemical cycles continuously 
supply energy to motor 
dyneins within flagellar axonemes \cite{Wemmer2004}.  Stepping action of individual motors is intrinsically stochastic 
\cite{Shingyoji1998}, and yet, beating can nevertheless persist, resilient against a cacophony of 
biochemical and background fluctuations. In assessing the fidelity or robustness of a 
biological oscillator, the stability and rhythmicity of its oscillations serve as prime indicators: one might identify 
pathological gaits of human walking from measures of 
cycle stability \cite{Bruijn2013}, determine the phase-dependent response of circadian clocks using 
external stimuli \cite{Mihalcescu2004}, or 
infer the health of a human heart from variability of inter-beat intervals \cite{Ashkenazy2001,Ivanov1996}. 
While periodic oscillations of beating flagella are correlated with a cell's responses and 
sensitivity to its environment, study of these features remains inchoate
\cite{Polin2009,Goldstein2009,Leptos2013,Wan2014,Geyer2013,Ma2014}. 

\begin{figure}[b]
\centering
\includegraphics[width=0.45\textwidth]{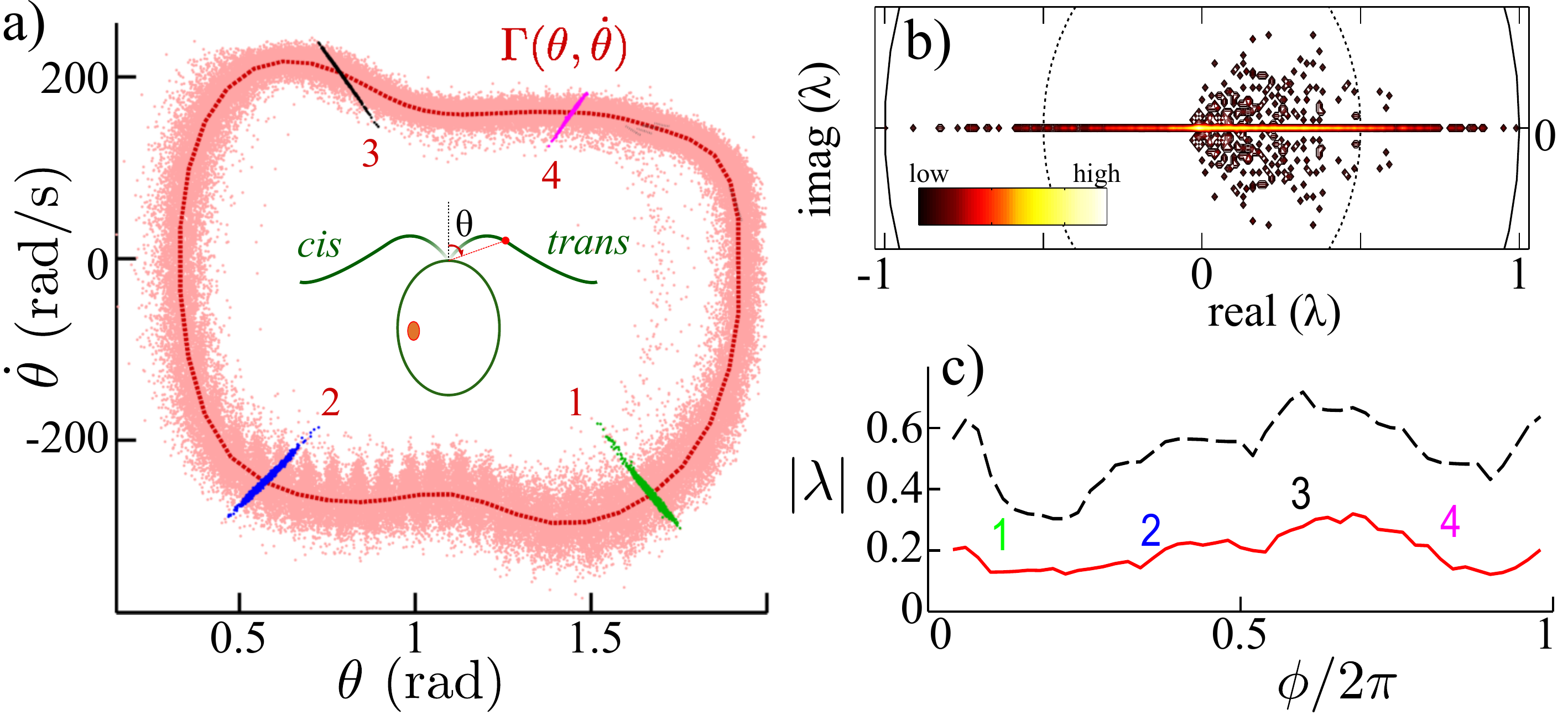}
\caption{(color online) Noisy flagellar limit cycles.  
a) Trajectories in ($\theta,~\dot\theta$) space
at fixed arclength ($=(2/7)$ of the total flagellum length). 
Four Poincar\'e sections are highlighted.
For the population (n $=48$): b) shows an accumulated density map of Floquet multipliers $\{\lambda\}$ computed at different phases, while in c) the distribution of $|\lambda|$ is characterized by its mean (solid line) and $95$th percentile (dashed line).}
\label{fig:fig1}
\end{figure}

\begin{figure*}[t]
\centering
\includegraphics[width=0.84\textwidth]{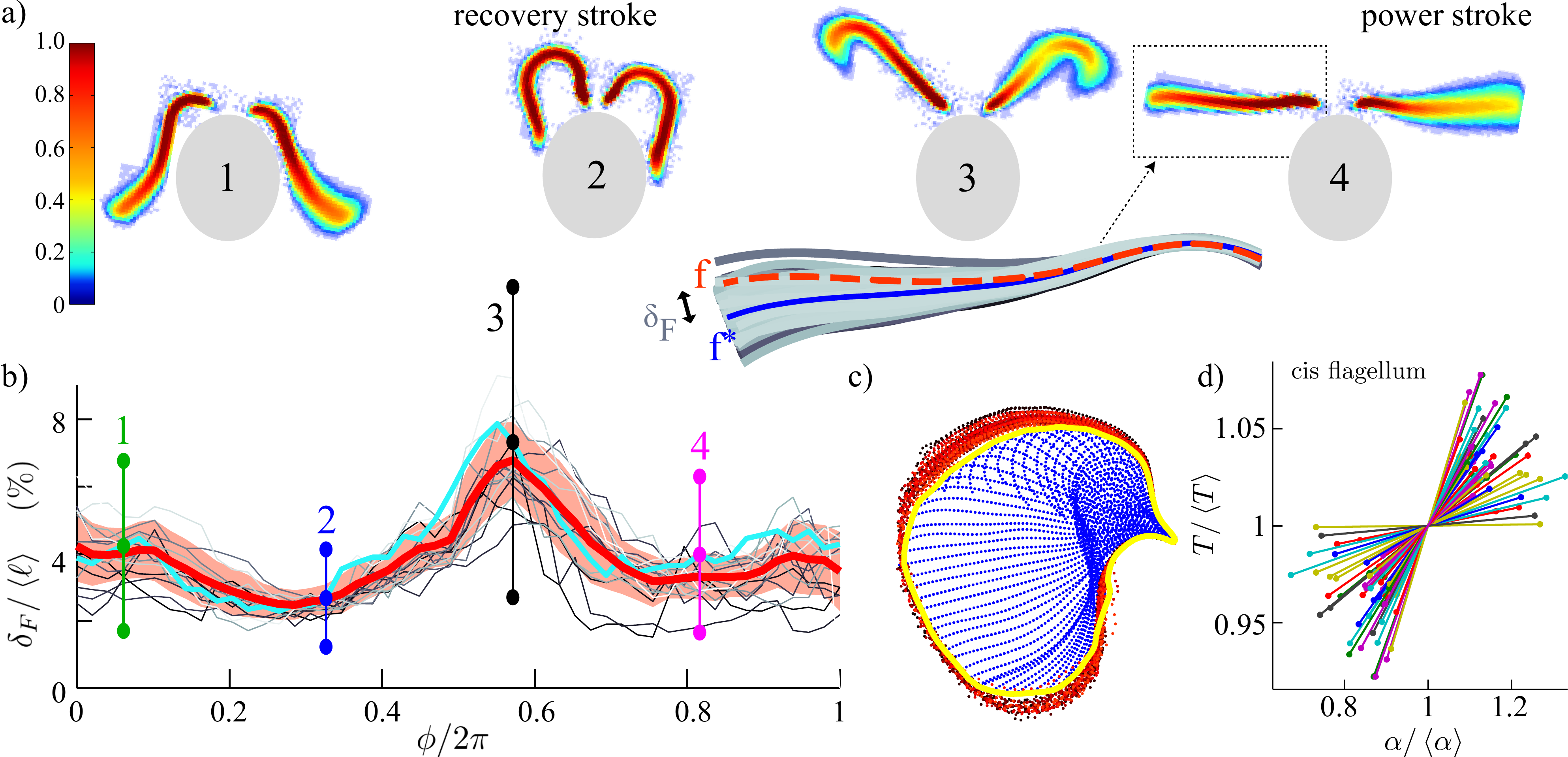}
\caption{\small (color online) Noisy flagellar waveforms. 
a) Overlaid waveforms at phases $1$--$4$ (Fig.~$1$a), colored by recurrence.   
Iso-phase waveforms $\mb{f}$ coalesce in a band about an average shape ${\mb{f}}^*$.			
b) Length-normalized Fr\'echet distance $\delta_F(\mb{f},{\mb{f}}^*)/\ell$ computed for multiple cells, showing 
phase-dependent noise. (Cyan: average over ${\cal O}(10^3)$ beat cycles for a single cell; error bars: one s.d. from mean. 
Red: a multicell average; shading: one s.d. from mean.) 
c) Discretized points (blue) along a flagellum define an area \textit{per beat} via an alpha-shape (yellow), which fluctuates over successive beat cycles (red). 
d) Per-beat area $\alpha$ and per-beat period $T$ are strongly correlated. Individual lines summarize the per-cell noisy scatter.
\label{fig:fig2}}
\end{figure*}

Here, drawing on data from a large population of cells ($\sim\!\!100$), we examine fluctuations in beating due 
to perturbations that are a) \textit{continuous}, or b) \textit{transient}.
Case (a) encompasses contributions from sources over which the experimenter has little control: background thermal noise, intracellular biochemical processes \cite{Sakakibara1999}, or even photon irradiance \cite{Hegemann1988}. 
We find that flagellar dynamics are stable to these weak fluctuations, but waveform noise displays an intriguing phase dependence, or {\it rhythmicity}.
Beat-to-beat intervals form timeseries exhibiting fractal structure, and successive 
beats may remain correlated for many seconds, even displaying oscillatory correlation ({\it recurrence}). 
Yet in cells placed under physiological stress, measured timeseries signal much more erratic and variable flagellar 
beating dynamics. 
For (b), we inject fluid impulses near a beating flagellum and examine the post-perturbation relaxation 
to the stable limit cycle of the breaststroke.
This {\it recovery} from beating disruption is a crucial property of viable cilia and flagella.
     
To permit long-time, in-focus visualization of flagellar dynamics, wildtype cells 
(strains CC$124$ and CC$125$, \textit{Chlamydomonas} Center) were individually caught and fixed by micropipette micromanipulation (Patchstar, Scientifica, UK) with gentle suction \cite{Polin2009,Wan2014}. 
High-speed images (SA$3$ Fastcam, Photron, USA and Phantom V311, Vision Research, USA) of beating flagella were captured at $2000-3000$ frames/s -- at least one order of magnitude above the natural beat frequency. 
Continuous recordings ($1-10$ minutes) were taken for each cell, from which $\sim\!\! 1-5\times10^3$ contiguous 
beat cycles could be extracted.  
Recordings were conducted under conditions that appropriately mimic a cell's natural daytime habitat, namely white light illumination (halogen lamp), and hence some phototactic response is expected \cite{Ruffer1998}. 
Pixel coordinates that track the flagellum in each frame were converted to spline fits, and used to generate timeseries. 
		
Automated waveform tracking gives unprecedented spatio-temporal resolution \cite{Wan2014},
which over thousands of cycles allows determination of the spatial reproducibility of beating.
Relative to a reference axis, angles $\theta(t)$ traced by a point at fixed arclength \cite{WanNB1} (Fig.~\ref{fig:fig1}a) 
are projections of the multidimensional dynamics. 
The point cloud ($\theta,\dot{\theta}$) maps the attracting region around a limit cycle 
$\Gamma$ -- approximated numerically. 
Progression through each cycle was charted by associating the $2D$ flagellum centerline $\mb{f}(t_i)$ at time $t_i$ with a uniformly-rotating phase $\phi=\omega_0t$ defined from the polar angle $\varphi=\tan^{-1}(\theta-\left\langle\theta\right\rangle)/(\dot{\theta}-\dot{\left\langle\theta\right\rangle})$  using the transformation $\phi=\omega_0\int(d\varphi/dt)^{-1}\,d\varphi$.
Ratio distributions are approximated using Fourier series \cite{Wan2014}. 
Trajectory crossings $C=\{\mb{x}_n:\phi({\cal P}^n(\mb{x}_n))=\phi_0,~n = 1,2,3,\ldots\}$ at fixed $\phi=\phi_0$ correspond to iterations of a Poincar\'e return map ${\cal P}$.
We computed for each cell and $50$ subdivisions of $\left[0,2\pi\right]$, eigenvalues of the Jacobian matrix of derivatives ${\cal J}={D\cal P}|_{\mb{x}^*}$ taking $\mb{x}^*=\left\langle\mb{x}\right\rangle_{\mb{x}\in C}$, and 
fitting to the bilinear model $(\mb{x}_{n+1}-\mb{x}^*)={\cal J}(\mb{x}_{n}-\mb{x}^*)$.
The distribution of computed eigenvalues (Fig.~$1$b) is particularly dense on the real line.
All eigenvalues have magnitude less than unity, fulfilling our intuition that limit cycles corresponding to the breaststroke gait are stable.

To examine the phase-dependence in the noise suggested by Fig.~1c, we appeal to the full dimensionality of the waveforms.
The set $S_{k}=\{\mb{f}(t_j^k)\}$, where $\{j:\phi(t)|_{t=t_j^k}=\phi_k\}$ for phases $\phi_k=2\pi k/50$, $k=1,\cdots, 50$, 
groups periodic waveforms at equivalent phase (Fig.~\ref{fig:fig2}a).
We measure the dissimilarity between $\mb{f}\in S_k$ and an average waveform ${\mb{f}}^*_k$ by a (discrete) {\it Fr\'echet distance} 
\begin{equation}
\delta_F(\mb{f},{\mb{f}}^*_k):=\min_{\cal U}\max_{(\mb{p},\mb{q})\in {\cal U}}||\mb{p}-\mb{q}||,
\end{equation}
with tracked waveforms approximated by polygonal curves corresponding to ordered vertices $\sigma(\mb{f})=(\mb{p}_1,\cdots,\mb{p}_m)$ and $\pi(\mb{f}^*_k)=(\mb{q}_1,\cdots,\mb{q}_n)$ ($m,n\in\mathbb{Z}$), and 
\[
{\cal U} = \left\{(\mb{p}_{u_i},\mb{q}_{v_j})\in\sigma(\mb{f})\times\pi(\mb{f}^*_k) \quad (i,j=1,\cdots, J)\right\}, 
\]
comprising $J$ pairs of vertices which are complete (for every $\mb{p}\in\sigma$ there exists $i,j$ with $(\mb{p}_i,\mb{q}_j)\in{\cal U}$ and $\mb{p}=\mb{p}_i$; similarly for $\mb{q}\in\pi$) and ordered ($u_{i+1}=u_{i}$ or $u_{i+1}=1+u_{i}$, and $v_{j+1}=v_{j}$ or $v_{j+1}=1+v_{j}$).
The computation is performed recursively, in ${\cal O}(mn)$ time \cite{Eiter1994}.
At each phase $\delta_F$ gauges \textit{waveform noise} in the periodic formation of the flagellum shape (Fig.~\ref{fig:fig2}b),
and is minimized during recovery strokes ($\lesssim 1.7\%$ -- a value 
comparable to measurement noise), and maximized at the transitions between power and recovery strokes ($\gtrsim 10.8\%$).

\begin{figure*}[t]
\centering
\includegraphics[width=0.85\textwidth]{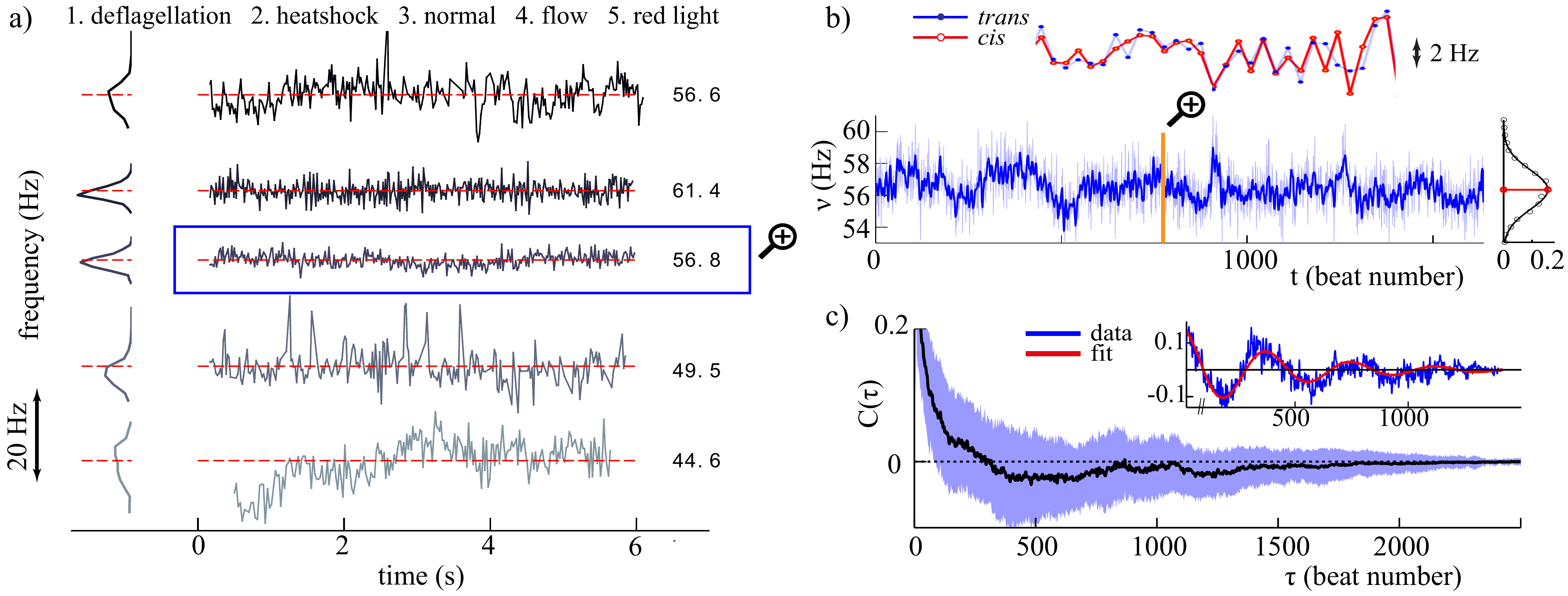}
\caption{(color online) Correlations in flagellar beating.	
a) Signatures of interbeat frequencies $\nu(t)$ in the flagella of cells, observed in a number of scenarios ($1-5$, see main text).
(Timeseries have been displaced vertically, with mean frequencies as labelled.)
b) Long-range fluctuations in $\nu(t)$ are observed. For a control cell (case $3$), the filtered signal is superimposed with the raw data, and its pdf fit to a Gaussian. \textit{Cis} and \textit{trans} flagella of the same pair exhibit perfect frequency-locking, highlighting the accuracy of the measurement technique.
c) Decay of autocorrelation in interbeat intervals $b(t):=1/\nu(t)$, showing the population average (solid line) and one s.d. from the mean (shaded). Inset: parametric fit to a sample $C(\tau)$.
\label{fig:fig3}}
\end{figure*}
In a classic eukaryotic flagellum, beating emerges from periodic, selective activation of motor dyneins that crosslink internal filaments \cite{Wemmer2004}.
At putative switch-points between power and recovery strokes \cite{Lindemann2010}, geometrically opposed groups of dyneins detach on one side and reattach at the other to their respective microtubule tracks, until beating direction is reversed.
Thus, high waveform noise correlates with a large number of activated dyneins.

The timing of flagellar strokes is determined by the microscale action of dyneins, which in turn governs the frequency and amplitude of the beat.
Here we partition flagellar positions by phase, averaging two different Poincar\'e sections to obtain the instantaneous period $T_n$ and frequency $\nu_n=1/T_n$, indexed by beat number $n$.
From the data, we approximated the $n$th-cycle beat envelope by its \textit{alpha-shape} \cite{alpha1983}, which generalizes the concept of a convex hull (Fig.~\ref{fig:fig2}c).
Accuracy in the computed alpha-shape area $\alpha_n$ is defined up to discs of radii $5$~pixels$~\!\approx 1.11$ $\mu$m.
We find $T_n$ and $\alpha_n$ to be strongly correlated. 
Denoting by $\left\langle \cdot\right\rangle$ an average over beat cycles, and plotting ${\Pi}=T_n/\left\langle T_n\right\rangle$ vs ${A}=\alpha_n/\left\langle \alpha_n\right\rangle$ reveals directional but very noisy scatter.
A similar correlation has been found independently \cite{Mai2014}.
To characterize this directionality we compute the matrix 
\begin{equation}
\text{Cov}\left[\Pi,A\right]=\begin{pmatrix}
\left\langle{\cal T}{\cal T}\right\rangle&\left\langle{\cal T}{\cal A}\right\rangle\\ 
\left\langle{\cal A}{\cal T}\right\rangle&\left\langle{\cal A}{\cal A}\right\rangle
\end{pmatrix}~, 
\end{equation}
where ${\cal T}=\Pi-\left\langle \Pi\right\rangle$ and ${\cal A}=A-\left\langle A\right\rangle$.
From the timeseries for each cell $i$ we estimate ${\cal T}/{\cal A}$ by $\gamma=\tan^{-1}(v_2/v_1)$ where $(v_1,v_2)$ is the principal eigenvector direction (Fig.~\ref{fig:fig2}d).
We find $\overbar{\gamma}\sim(0.264\pm 0.146)$~rad where the bar denotes an ensemble average over multiple cells, and correspondingly a dimensional ratio of increments $\overbar{r}\approx39.7\pm 31.0~\mu$m$^2$/ms, where $r=\left\langle \alpha_n\right\rangle/\left\langle T_n\right\rangle\times 1/\tan(\gamma)$. 
Assuming a flagellum ``wingspan'' of $10~\mu$m during the power stroke this is equivalent to a velocity scale $\delta \ell/\delta T\sim4~\mu$m/ms, for an effective amplitude $\ell$.
A rod-like flagellum of length $\ell$ produces a motive force $F\sim\eta T^{-1}\ell^2$ and power density $P/\ell$ where $P\sim\eta T^{-2}\ell^3$ (where $\eta$ is the medium viscosity); that amplitude and frequency are inversely correlated suggests constancy of force and/or power production by axonemal motors, and is often assumed without proof in certain bead-on-spring models of beating cilia.
Fundamentally, hydrodynamic synchronization in coupled ciliary arrays also necessitates that (within a physiologically-relevant regime) decrease in beat frequency accompanies increase in amplitude \cite{Niedermayer2008}, such as we have demonstrated here.

The association of oscillatory dynamics with a well-defined frequency does not \textit{a priori} imply stability.
Stable flagellar beating, as we have now established for the canonical \textit{Chlamydomonas} breaststroke, does not generalize to all flagellate species nor to \textit{Chlamydomonas} cells that are physiologically ``abnormal''.
For instance the timeseries $\nu(t)$  in Fig.~\ref{fig:fig3}a are representative of flagellar beat frequency fluctuations in a number of scenarios of interest.
In case $1$ we initiated complex calcium fluctuations and repair processes in a cell \cite{Wheeler2008} by mechanical deflagellation of one flagellum; $\nu(t)$ was then measured for the remaining flagellum, which continues to beat as the amputated flagellum is regrown within $1\sim 2$ hours.
A heatshock treatment was used in case $2$ to disrupt enzymatic pathways \cite{Schulz-Raffelt2007}, in which cell cultures were immersed in a $35^\circ$ water bath for $10$ minutes prior to experimentation.
Case $3$ is a control cell.
Cells in case $4$ were subject to a frontally-directed flow, controlled by a syringe pump (PHD$2000$, Harvard Apparatus).
Filtering the illumination light ($>620$ nm filter) leads to persistent light-adaptation processes and frequency drift \cite{Leptos2013}; this is case $5$.
Noisy flagellar dynamics are thus a directly quantifiable measure of a cell's physiological state.

Even control cells (Fig.~\ref{fig:fig3}a, case $3$) are subject to persistent, weak environmental fluctuations that feedback-modulate flagellar beating.
Measured beat frequencies in the two flagella of a given cell agree with remarkable precision (Fig.~\ref{fig:fig3}b, inset).
From timeseries $b(t)$ of interbeat intervals we construct the statistic $C(\tau) = \langle b(t+\tau)b(t)-\langle b\rangle^2\rangle$, where $\left\langle\cdot\right\rangle$ denotes a time average. 
The decay of $C(\tau)$ was found to be unexpectedly slow, and in many cases even oscillatory (Fig. \ref{fig:fig3}c) -- suggestive of an underlying periodic process with noise. 
Let $b(t)=b_0(1+\beta(t))\cos(\omega_0 t+\phi(t))$, where $b_0$ and $\omega_0$ are the averaged 
amplitude and frequency of oscillations, and $\beta(t)$, $\phi(t)$ are independent functions respectively characterizing phase and amplitude noise.
We assume that $\beta(t)$ is stationary, and that $\phi(t)$ is a Brownian motion with $\left\langle\phi(t)\right\rangle=0$ 
and $\left\langle\phi(t)^2\right\rangle=Dt$.
The autocorrelation is
\begin{equation}
\tilde{C}(\tau)=\frac{b_0^2}{2}[1+C_b(\tau)]{\rm e}^{-D|\tau|}\cos(\omega_0 \tau)~,\label{eq:autocorr}
\end{equation}
where $C_b$ is the covariance of $\beta(t)$.
For a sample cell we fit $C(\tau)$ using (\ref{eq:autocorr}) with an empirical function $C_b(\tau)=\beta e^{-|\tau|/\xi}$ (Fig.~\ref{fig:fig3}c, inset), yielding $b_0=0.157$, $D=0.002$, $\omega_0=0.016$, $\beta=9.928$ and $\xi=1.85$.
In particular we find a timescale for the periodicity of slow oscillations: $2\pi/\omega_0=392$ beats, or $6.01$~s.
Sampled over $65$ cells, the average form of $C(\tau)$ takes $\sim\!250$ beats for the correlation to reverse sign, and 
persists over $\sim\! 1000$ beats, or $\sim\! 15$~s.

Our $b(t)$ timeseries possess fractal structure, and is correlated across multiple scales.
In the first instance we can derive a scalar measure $\alpha$ via a detrended fluctuation analysis (DFA) to characterize individual timeseries \cite{Peng1994}, as follows.
Construct first the integrated signal $B(t_j)=\sum_{i=1}^j(b(t_i)-\left\langle b\right\rangle)$,~$(1\leq j\leq L)$.
Then for $K$ sections $\{I_i:=[t_i,t_{i+1}],~t_i=iL/K,~i = 1,2,\cdots,K-1\}$ each of size $N=L/K$, the local trend in $B$ is computed at the $i$th section (let $B_N(t_i)$ be the lsq linear fit to data points $B(t_i\in I_i)$).
The fluctuation
\begin{equation}
 F(N)=\sqrt{\frac{1}{L}\sum_{i=1}^L(B(t_i)-B_N(t_i))^2}~,
\end{equation}
is computed at multiple scales and a power-law scaling $F(N)\sim N^\alpha$ is obtained.
We calculated $\alpha = 0.83\pm 0.10$ (for $67$ cells, ${\cal O}(10^3)$ successive beats each).
This persistent positive correlation is lost upon randomly permuting $b(t)$ (each time averaging over $10$ shuffles), which yields $\alpha = 0.48\pm0.03$, consistent with white noise.

The frequency (and hence synchrony) of flagellar beating is controlled at a biomolecular level by calcium \cite{Bessen1980,DiPetrillo2010}.
Previously we found that the flagella of freeswimming {\it Chlamydomonas} switch stochastically \cite{Polin2009} between synchronous and 
asynchronous beating (drifts) on a timescale of $\sim\!10$~s, and suggested this may be due to calcium fluctuations which affect {\it cis} 
and {\it trans} flagella differentially \cite{Wan2014}. 
Our present discovery of slow oscillations in flagellar beat frequency might then relate these transitions in beating modes  
to stochastic crossings of a putative calcium threshold.
Fluctuations in cytosolic calcium of ${\cal O}(s)$ have been measured \textit{in vivo}, in \textit{Chlamydomonas} cells ballistically-loaded 
with calcium dyes \cite{Wheeler2008}.

\begin{figure}[t]
\centering
\includegraphics[width=0.4\textwidth]{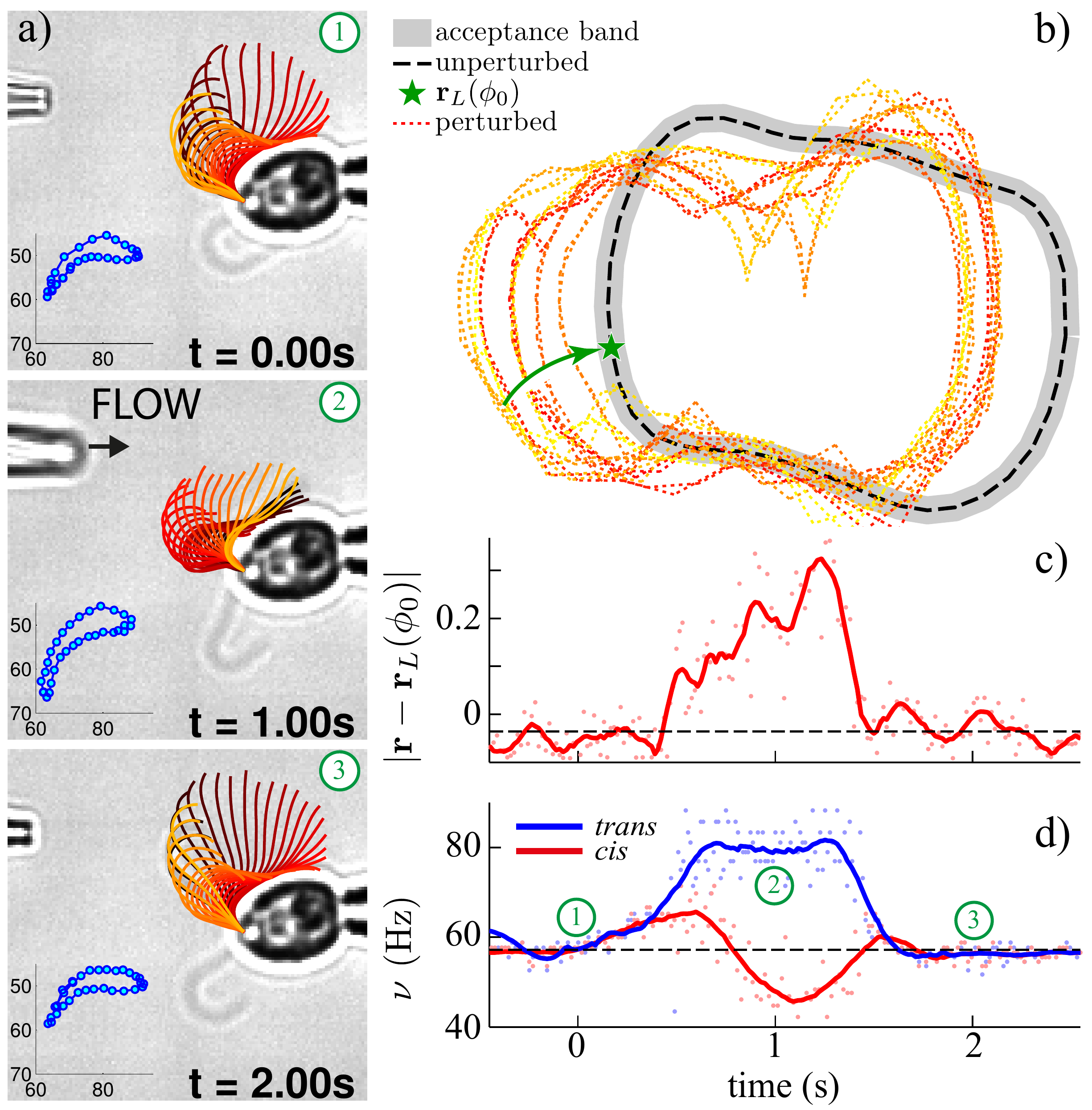}
\caption{(color online) Stability to perturbations.
a) Fluid is injected from a $2$nd pipette (arrow). Waveform sequences for the \textit{cis} flagellum only are shown ($1$-$3$). Insets: $x$-$y$ coordinates of a reference point at fixed arclength. 
b) Trajectories veer off the pre-perturbation limit cycle during one perturbation event.
This deviation is sampled at fixed phase as a function of time (c), which accompanies marked changes in the beat frequencies of both flagella (d).
}
\label{fig:fig4}
\end{figure}

Sudden elevations in intracellular calcium can be triggered either by activation of photoreceptors in the eyespot \cite{Foster1980} or of mechanosenstive channels in the membrane \cite{Fujiu2009}, leading to altered flagellar beating.
By perturbing a beating flagellum with manually induced pulses of fluid from a $2$nd pipette (which delivers $\sim\!100$~pN forces according to PIV measurements), we can compute the attractor strength $\sigma$ of flagellar oscillations (Fig~\ref{fig:fig4}a, 1-3).
Limit cycles and phases are defined from tracked waveforms as previously (Fig.~\ref{fig:fig1}).
The pre-perturbation cycle $\mb{r}_L$ was chosen as reference. 
If perturbed trajectories $\mb{r}(t)$ evolving in time $t$ contract linearly towards the stable attractor, then at a representative phase $\phi=\phi_0$, 
\begin{equation}
\sigma = \frac{1}{\tau_\infty-\tau_0}\ln\frac{|{\mb{r}(\tau_\infty; \phi_0)-\mb{r}_L(\phi_0)}|}{|{\mb{r}(\tau_0; \phi_0)-\mb{r}_L(\phi_0)}|}~,
\end{equation}
where $\tau_0$ is chosen at maximum deviation, and $\tau_\infty$ when $\mb{r}$ first returns (and remains) within an acceptance band about $\mb{r}_L$.
Averaging multiple experiments, we find $\sigma\sim 2.94\pm1.72$ s$^{-1}$ or $\sigma^{-1}\sim 20.4$ beats.
Thus normal flagellar beating can readily (and in characteristic time) recover from moderate hydrodynamic disturbances which mimic that which microalgae encounter in their native habitats.
If local perturbation of one flagellum transiently elevates intracellular calcium, the observation of altered beating of \textit{both} flagella in a coupled pair (Fig.~4d) is consistent with differential \textit{cis}-\textit{trans} flagellar calcium response, or dominance \cite{Wan2014}.
This rapid loss of biflagellar synchrony implicates internal biochemical control of normal breaststroke coordination (Fig.~3b,inset); in contrast the beating of flagella belonging to \textit{different} cells can be synchronized solely by the hydrodynamics \cite{Brumley2014}.

Through dynamic high-resolution tracking, the rhythmicity of eukaryotic flagellar oscillations was revealed and the nature of flagellum noise explored.
We demonstrated significant spatio-temporal correlation in the beating dynamics, and suggested that while variations on timescales of beat-cycles 
may be due to intrinsic motor stochasticity, long-range correlations in beat frequency may be signatures of \textit{in vivo} biochemical signalling 
via second messengers such as calcium \cite{calcium2011}.
Indeed calcium governs ciliary beating in many different organisms \cite{Evans1999,Schmid2011,Salathe1999}; oscillatory calcium dynamics would 
vastly improve specificity, allowing signals to integrate without sustained rise.
It would be interesting to examine the noise spectrum of beating in artificial or reconstituted flagella, where feedback-regulation would take on a 
very different form. 

We thank M. Polin, K.C. Leptos, and P. Holmes for discussions. 
Financial support is acknowledged from the EPSRC, ERC Advanced Investigator Grant 247333, and a Senior Investigator 
Award from the Wellcome Trust.

\bibliographystyle{apsrev4-1}
\bibliography{wan_refs}

\begin{thebibliography}{33}
\bibitem [{Wan()}]{WanNB1}%
  \BibitemOpen
  \href@noop {} {}\bibinfo {note} {These results were found to hold
  independently of this choice of reference arclength.}\BibitemShut {Stop}%
\end{thebibliography}
\end{document}